\documentclass[12pt]{article}

\usepackage[totalheight = 23cm, totalwidth = 17cm]{geometry}
\usepackage{amssymb,amsmath,amsfonts,amsbsy,graphicx,bm}

\usepackage{color,cancel,ulem}

\newcommand{\dis}[1]{\begin{equation}\begin{split}#1\end{split}\end{equation}}

\begin{document}

\begin{titlepage}

\begin{center}

{\LARGE \bf 
(In)stability of de Sitter vacuum in light of distance conjecture and emergence proposal 
}

\vskip 1.0cm

{\large
Min-Seok Seo$^{a}$ 
}

\vskip 0.5cm

{\it
$^{a}$Department of Physics Education, Korea National University of Education,
\\ 
Cheongju 28173, Republic of Korea
}

\vskip 1.2cm

\end{center}

\begin{abstract}

 The distance conjecture claims that   as the modulus traverses along the trans-Planckian geodesic distance, the effective field theory becomes invalid by a descent of a tower of states  from UV.
 Moreover, according to the recent (strong version of) emergence proposal, the kinetic term of the modulus is entirely generated by the wavefunction renormalization in which a tower of states are integrated out.
 Assuming these two conjectures, we explore the role of a tower of states coupled to the modulus in (in)stability of the de Sitter (dS) vacuum by studying the one-loop effective potential generated by a tower of states.
 We find that a fermionic tower of states makes the effective potential more or less consistent with the dS swampland conjecture : either the   slope or the curvature of the potential is sizeable.
 In contrast, the effective potential generated by a bosonic tower of states seems to allow the stable dS vacuum.
 Therefore, in order to argue the instability of the dS vacuum, the additional ingredient like supersymmetry breaking needs to be taken into account.

\end{abstract}

\end{titlepage}

\newpage

\section{Introduction}

  Difficulties in realizing a metastable de Sitter (dS) vacuum in the context of string theory have raised the suspicion that the background geometry with the positive cosmological constant is destabilized by quantum gravity effects (for recent reviews, see, e.g., \cite{Berglund:2022qsb,  Cicoli:2023opf}). 
  This concern was formulated as the `dS swampland conjecture' \cite{Obied:2018sgi} (see also \cite{Danielsson:2018ztv, Cicoli:2018kdo}), the refined version of which \cite{Andriot:2018wzk, Andriot:2018mav, Garg:2018reu, Ooguri:2018wrx} states that the effective scalar potential consistent with quantum gravity satisfies either
 \dis{m_{\rm Pl}\frac{|\nabla V|}{V}\sim {\cal O}(1)\quad\quad{\rm or}\quad\quad
 {\rm min} \Big(m_{\rm Pl}^2\frac{\nabla_i\nabla_j V}{V}\Big)\leq -{\cal O}(1).} 
 As argued in \cite{Ooguri:2018wrx}, the dS swampland conjecture can be supported by the distance conjecture \cite{Ooguri:2006in}, together with the covariant entropy bound \cite{Bousso:1999xy}.
The distance conjecture predicts that   the infinite distance limit of the  moduli space corresponds to a particular corner of the landscape at which the mass scale of a tower of states becomes extremely light. 
Then the low energy effective field theory (EFT) description becomes invalid at this corner as an infinite tower of states descends from UV.
In particular, the covariant entropy bound of dS space can be violated by the rapid increase in the number of low energy degrees of freedom.

 In order to quantify the validity of the EFT in light of the distance conjecture, we first need to specify an appropriate UV cutoff scale above which the EFT is no longer reliable.
  Whereas the Planck scale $m_{\rm Pl}$ is a natural UV cutoff in quantum gravity, if we restrict our attention to the EFT in which gravity is weakly coupled, the lower scale called the species scale can be used as the UV cutoff.
  Here  the species scale  $\Lambda_{\rm sp}$ is defined by the scale above which   the strength of the gravitational coupling $N_{\rm sp} (\Lambda_{\rm sp}/m_{\rm Pl})^2$ becomes larger than  ${\cal O}(1)$, where  $N_{\rm sp}$ is the number of light degrees of freedom below $\Lambda_{\rm sp}$ \cite{Dvali:2007hz, Dvali:2007wp}.
 According to the distance conjecture,  as the modulus traverses along the trans-Planckian geodesic distance, at least one of tower mass scales becomes extremely light, hence $N_{\rm sp}$ increases rapidly.
 This results in the rapid decrease in  $\Lambda_{\rm sp}$, and the EFT becomes invalid when $\Lambda_{\rm sp}$  is even smaller than the typical mass scale  of the EFT.
 For the EFT in the background with the positive cosmological constant, the characteristic mass scale is given by the Hubble parameter $H$, the inverse of the horizon radius.
 Then the valid EFT description requires that $\Lambda_{\rm sp}$ must be larger than $H$.

  Meanwhile, the geodesic distance of the  modulus can be connected to  the tower mass scale through the  coupling between the modulus and a tower of states.
From this we expect  that parameters describing the IR dynamics of the modulus  get renormalized by the loop contributions from a tower of states below $\Lambda_{\rm sp}$ significantly.
  Regarding the wavefunction renormalization, the role of a tower of states is conjectured under the name of the `emergence proposal' \cite{Harlow:2015lma, Heidenreich:2017sim, Grimm:2018ohb, Heidenreich:2018kpg, Castellano:2022bvr} (for recent relevant studies, see, e.g., \cite{Corvilain:2018lgw, Blumenhagen:2019qcg, Blumenhagen:2019vgj, Marchesano:2022axe, Castellano:2023qhp, Blumenhagen:2023yws}).
  The weak and conservative version of the proposal predicts the existence of a tower of states which induces quantum corrections to the kinetic term matching the tree level singular behavior. 
   On the other hand, the strong version of the emergence proposal suggests that all light particles in the weakly coupled regime in fact do not have the kinetic terms in UV : the kinetic terms of the low energy degrees of freedom are an IR effect generated by integrating out   towers of light states below $\Lambda_{\rm sp}$.
  In other words, the modulus can have the kinetic term only after a tower of states coupled to the modulus is integrated out to generated the wavefunction renormalization.

 Moreover, as the tower mass scale decreases along the trajectory of the modulus, the loop contribution of a tower of states to  the vacuum energy  changes, from which  the effective potential for the modulus can be obtained \cite{Coleman:1973jx}.
 Then the role of the distance conjecture in the (in)stability of dS space can be studied in the field theoretic language by investigating the behavior of the effective potential.
 For this purpose, we suppose that at   the initial value of the modulus   the vacuum energy density is given by the positive value, $3m_{\rm Pl}^2 H^2$, and $\Lambda_{\rm sp}$ is much larger than the characteristic scale of the EFT, namely, $H$.
 If the potential decreases rapidly as the modulus moves away from the initial value, we can say that the de Sitter vacuum is destabilized by the decrease in the tower mass scale.  
  In contrast, if the potential is stabilized before the vacuum energy becomes negative, the dS vacuum is not destabilized by a tower of states at least at the field theoretic level.

 In this article, we assume the distance conjecture and the strong version of the emergence proposal such that both the effective potential and the kinetic term of the modulus are entirely generated by one-loop contributions of a tower of states.  
 Then we  explore the (in)stability of dS space reflected in the  effective potential.   
 In Section \ref{sec:waveren}, we first review how the one-loop wavefunction renormalization of the modulus is generated by integrating out  the bosonic and fermionic towers of states.
 When the strong version of the emergence proposal is assumed, it turns out that the canonically normalized modulus is given by the exponent of   the tower mass scale.  
 In Section \ref{sec:Effpot}, the one-loop effective potential of the modulus induced by a tower  of states is calculated.
Since  the tower mass scale is typically heavier than $H$  the leading term of the effective potential in the background with the positive cosmological constant is the same as that  in the flat spacetime background.
Nevertheless, we work on the curved background from beginning, which may be useful in the future work considering the subleading terms of the effective potential.
 Since the closed fermion loop contains the extra minus sign, the behavior of the effective potential generated by a fermionic tower of states is different from that generated by a bosonic one.
 Discussion in Section \ref{sec:dSstab} shows that the effective potential generated by a fermionic tower of states is well consistent with the refined dS swampland conjecture.
 In contrast, the effective potential generated by a bosonic tower of states seems to allow the stable dS vacuum by the appropriate renormalization.
 This suggests that to argue the instability of dS space, we need another reason.
 After discussing the possible additional ingredient for the instability of dS space, we conclude.

\section{Wavefunction renormalization of the modulus}
\label{sec:waveren}

 In the presence of a tower of states with the mass scale  $\Delta m$, the species scale above which gravity becomes strong satisfies
 \dis{\Lambda_{\rm sp}=\frac{m_{\rm Pl}}{\sqrt{N_{\rm sp}}},}
 where $N_{\rm sp}=\Lambda_{\rm sp}/\Delta m$ is the number of  states in a tower with masses $m_n=n \Delta m$ ($n \in \mathbb{Z}$) below $\Lambda_{\rm sp}$.
 In terms of $\Delta m$, $N_{\rm sp}$ and $\Lambda_{\rm sp}$ are given by
 \dis{N_{\rm sp}=\Big(\frac{m_{\rm Pl}}{\Delta m}\Big)^{2/3},\quad
 \Lambda_{\rm sp}=(m_{\rm Pl}^2 \Delta m)^{1/3},\label{eq:NspLsp}}
 respectively.
 Here $\Lambda_{\rm sp}$ is assumed to be much larger than $\Delta m$ such that $N_{\rm sp} \gg 1$ is satisfied.
 We will consider the loop contributions of a tower of states below $\Lambda_{\rm sp}$ to the IR dynamics of the modulus $\phi$ when the tower mass scale $\Delta m$ depends on  the field value of $\phi$.
 
 \begin{figure}[!t]
  \begin{center}
   \includegraphics[width=0.4\textwidth]{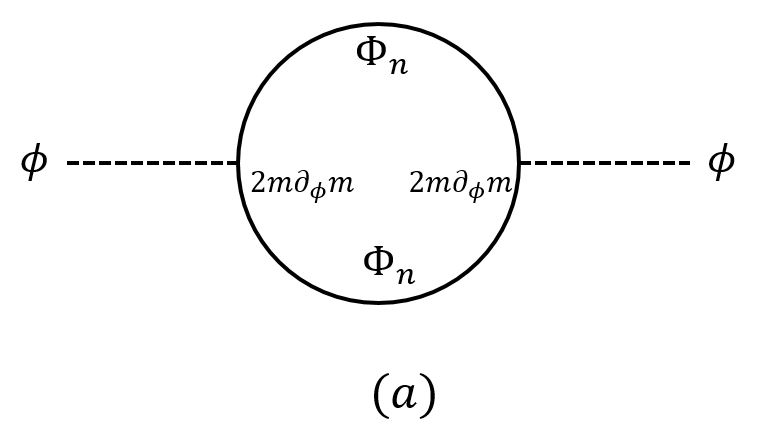}
   \includegraphics[width=0.4\textwidth]{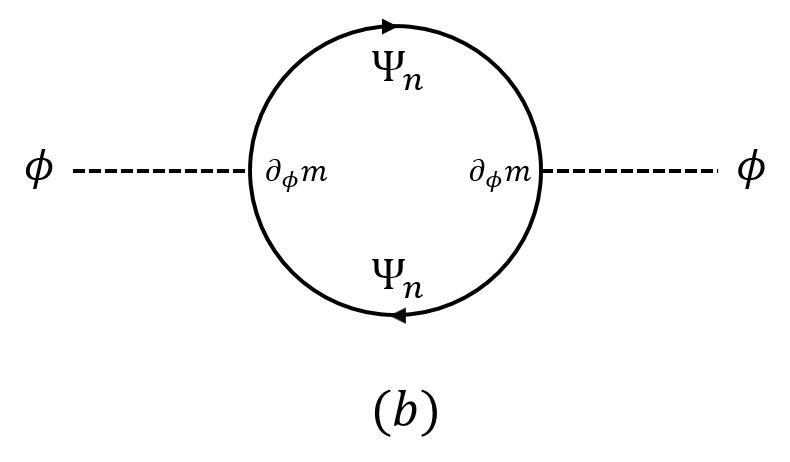}
  \end{center}
 \caption{One-loop diagrams for the wavefunction renormalization of the modulus $\phi$.
 Contributions from a bosonic and fermionic tower of states are depicted in  $(a)$ and $(b)$, respectively.
  }
\label{fig:oneloop}
\end{figure}

For a bosonic (more precisely, scalar) tower of states $\Phi_n$, the mass term can be expanded as
\dis{\frac12 m_n^2(\phi)\Phi_n^2 =\frac12\Big[m_n^2(\langle\phi\rangle)\Phi_n^2+2(m_n \partial_\phi m_n)\big|_{\langle\phi\rangle}\delta\phi \Phi_n^2+\cdots\Big], \label{eq:mexpand}}
where $\delta\phi=\phi-\langle\phi\rangle$ is the shifted field around the (classical) field value.
When $\Phi_n$ are integrated out, the second term in the RHS of \eqref{eq:mexpand} gives rise to $Z_{\phi\phi}$, the one-loop wavefunction  renormalization  of $\phi$ as depicted in  Figure \ref{fig:oneloop} $(a)$. 
By an explicit calculation, one obtains   \cite{Grimm:2018ohb} 
\dis{Z_{\phi\phi}&=\sum_{n=1}^{N_{\rm sp}}\frac{(\partial_\phi m_n)^2}{4\pi^2}\Big(\frac{2\pi}{3\sqrt3}-1\Big)=  \frac{(\partial_\phi \Delta m)^2}{4\pi^2}\Big(\frac{2\pi}{3\sqrt3}-1\Big)\sum_{n=1}^{N_{\rm sp}} n^2
\\
&\simeq \frac{(\partial_\phi \Delta m)^2}{4\pi^2}\frac{N_{\rm sp}^3}{15},}
where we use  $N_{\rm sp}\gg 1$ and $[(2\pi)/(3\sqrt3)]-1\simeq 1/5$.
From \eqref{eq:NspLsp}, this can be rewritten as 
\dis{Z_{\phi\phi}\simeq \Big(\frac{m_{\rm Pl}}{8\pi}\frac{\partial_\phi \Delta m}{\Delta m}\Big)^2.}

Meanwhile, for a (Dirac) fermionic tower of states $\Psi_n$, the expansion of the mass term leads to
\dis{m_n(\phi)\overline{\Psi}_n \Psi_n =m_n(\langle\phi\rangle)\overline{\Psi}_n \Psi_n +\partial_\phi m_n\big|_{\langle\phi\rangle}\delta\phi \overline{\Psi}_n \Psi_n+\cdots,}
from the second term of which we can consider the one-loop wavefunction renormalizaton  depicted in Figure \ref{fig:oneloop} $(b)$.
Summing up the contributions of $\Psi_n$ to the wavefunction renormalization, we obtain \cite{Grimm:2018ohb} 
\dis{Z_{\phi\phi} =&\sum_{n=1}^{N_{\rm sp}}\frac{(\partial_\phi m_n)^2}{4\pi^2}\log\Big(\frac{\Lambda_{\rm sp}^2}{m_n^2}\Big)=  \frac{(\partial_\phi \Delta m)^2}{4\pi^2} \sum_{n=1}^{N_{\rm sp}} n^2 \log\Big(\frac{N_{\rm sp}^2}{n^2}\Big)
\\
&\simeq \frac{(\partial_\phi \Delta m)^2}{4\pi^2}\frac{2 N_{\rm sp}^3}{9},}
where for the last expression, we approximated the summation with the integral, which is valid for $N_{\rm sp}\gg 1$.
From \eqref{eq:NspLsp}, this can be rewritten as 
\dis{Z_{\phi\phi}\simeq \Big(\frac{m_{\rm Pl}}{4\pi}\frac{\partial_\phi \Delta m}{\Delta m}\Big)^2.}

Above results show that if the kinetic term of the modulus $\phi$ in IR is emerged by integrating out a tower of states, it is given by
\dis{\frac12 Z_{\phi \phi}g^{\mu\nu}\partial_\mu \phi \partial_\nu \phi = \frac12 c^2\Big( m_{\rm Pl} \frac{\partial_\phi \Delta m}{\Delta m}\Big)^2 g^{\mu\nu}\partial_\mu \phi \partial_\nu \phi, }
where the loop factor $c$ depends on the statistics of a tower of states.
From this, we can redefine the modulus field such that  the kinetic term is written in the canonical form.
For this purpose we note that $\partial_\phi \Delta m$ is negative by the distance conjecture : the tower mass scale decreases rapidly as  the value of $\phi$ increases.
 Then the redefined modulus $\varphi$ gets larger as $\phi$ increases through the relation
 \dis{\partial_\mu \varphi =-c m_{\rm Pl} \frac{\partial_\phi \Delta m}{\Delta m}\partial_\mu \phi, }
 or equivalently,
 \dis{\varphi=c m_{\rm Pl}\log\Big(\frac{\Delta m(0)}{\Delta m(\phi)}\Big).\label{eq:varphi}}  
 That is, the canonically normalized modulus is given by the exponent of   the tower mass scale.

 Discussion so far visits the simplest case, where only the single tower of states contributes to the wavefunction renormalization.
 In the presence of several towers, the mass scales of which are labelled by $\Delta m_i$ ($i=1,2,\cdots$), the species scale $\Lambda_{\rm sp}$ and $N_i$, the number of states below $\Lambda_{\rm sp}$ for each tower, are given by
 \dis{\Lambda_{\rm sp}=\frac{m_{\rm Pl}}{(\sum_i N_i)^{1/2}},\quad\quad N_i=\frac{\Lambda_{\rm sp}}{\Delta m_i},}
 respectively.
 Then the wavefunction renormalization is given by the sum of  loops generated by all towers of states, 
 \dis{Z_{\phi\phi}&=\sum_i c_i^2 (\partial_\phi \Delta m_i)^2 N_{{\rm sp},i}^3 =\sum_i c_i^2 (\partial_\phi \Delta m_i)^2 \Big(\frac{\Lambda_{\rm sp}}{\Delta m_i}\Big)^3
 \\
 & \simeq \sum_i c_i^2 \Big(\frac{\partial_\phi \Delta m_i}{\Delta m_i}\Big)^2 \Big(\frac{\Lambda_{\rm sp}^3}{\Delta m_i}\Big).\label{eq:Zmany}}
 To obtain the canonically normalized field $\varphi$ as in  \eqref{eq:varphi}, we consider the following  specific cases.

 First, when one of tower mass scales, say, $\Delta m_1$ is extremely light compared to others, $N_1$ is much lager than $N_i$ ($i=2,3,\cdots$) and the species scale is approximated by $\Lambda_{\rm sp}\simeq m_{\rm Pl}/N_1^{1/2}$, almost independent of $N_i$ with $i\ne 1$. 
 If we assume that $m_{\rm Pl}\partial_\phi \Delta m_i/\Delta m_i \sim {\cal O}(1)$ for any tower mass scales, the wavefunction renormalization \eqref{eq:Zmany} is dominated by a tower  of states with the lightest mass scale :
 \dis{Z_{\phi\phi} \simeq c_1^2 \Big(\frac{\partial_\phi \Delta m_1}{\Delta m_1}\Big)^2  \frac{\Lambda_{\rm sp}^3}{\Delta m_1} = c_1^2 \Big( m_{\rm Pl} \frac{\partial_\phi \Delta m_1}{\Delta m_1}\Big)^2.}
 Then we can define the (approximate) canonically normalized field $\varphi$ by considering the lightest tower mass scale only.
  Meanwhile, if $n$ towers have the same  and the lightest tower mass scale  $\Delta m$, $N_i$  for these towers are given by the same value $N \equiv\Lambda_{\rm sp}/\Delta m$, satisfying $\sum_i N_i\simeq n N$.
 In this case, $\Lambda_{\rm sp}$ and $N$ can be written as
 \dis{\Lambda_{\rm sp}^3=\frac{1}{n}m_{\rm Pl}^2\Delta m,\quad\quad
 N^3=\frac{1}{n} \Big(\frac{m_{\rm Pl}}{\Delta m}\Big)^2, \label{eq:manylambdaN}}
 respectively.
 Then the wavefunction renormalization  \eqref{eq:Zmany} becomes
 \dis{{Z_{\phi\phi}\simeq \Big(\frac{1}n}\sum_i c_i^2\Big)\Big( m_{\rm Pl} \frac{\partial_\phi \Delta m }{\Delta m }\Big)^2,\label{eq:Ztotal}} 
 so we can define the canonically normalized field in the form of \eqref{eq:varphi} with the replacement of $c^2$ by $(\sum_i c_i^2)/n$.
  
 We note that a tower of states considered so far is the simplest type in which the mass is given by an integral multiple of the single tower mass scale.
  On the other hand, the mass of a tower of states can be associated with more than two  tower mass scales, as can be found in, for example, the compactification on more than two extra dimensions of comparable sizes  or the string compactification on the circle of radius close to the string length scale.
  In this case, the expressions for  $N_{\rm sp}$ and  $\Lambda_{\rm sp}$ given by \eqref{eq:NspLsp} are significantly modified as $N_{\rm sp}$ is given by the product, rather than the sum of the numbers of states associated with different tower mass scales.
  Moreover, even if these  tower mass scales are determined by the single modulus, a simple expression for the canonically normalized field like \eqref{eq:varphi} is in general not  defined.
  In Appendix \ref{app:multitower}, we discuss such differences in detail by considering the case of two tower mass scales as an example.

\section{One loop effective potential of the modulus}
\label{sec:Effpot}

As the tower mass scale decreases along the trajectory of $\phi$, their one-loop contributions to the vacuum energy also change, resulting in the deformation of  the (almost) flat potential depending on the field value of $\phi$. 
From this we obtain  the effective potential of $\phi$.
To see this, consider the metric of dS space in the flat coordinates,
\dis{ds^2=-dt^2+a(t)^2 d{\mathbf x}^2,\quad{\rm where}\quad a(t)=e^{Ht}.}
Then the mode expansion of the bosonic field $\Phi_n$ in a tower  is given by
\dis{\Phi_n=\int\frac{d^3 p}{(2\pi a(t))^{3/2}}e^{i \mathbf{p}\cdot\mathbf{x}}\Big(a_{\mathbf p}f_{\mathbf p}(t)+a_{-{\mathbf p}}^\dagger f^*_{\mathbf p}(t)\Big).}
Now let us rewrite $f_{\mathbf{p}}$ as
\dis{f_{\mathbf{p}} = \frac{1}{2W(t)}e^{-i \int^t W(t')dt'},}
such that $W(t)$ is well interpreted as the `energy' under the adiabatic expansion in which time derivatives like $(\dot{a}/a)^2=\ddot{a}/a=H^2$, as well as $(\dot{W}/W)^2$ and $\ddot{W}/W$ are smaller than $\mathbf{p}^2/a^2$ (see \cite{Parker:2009uva} and references therein).
Then the equation of motion $(\square+m_n(\phi)^2)\Phi_n=0$, or equivalently,
\dis{W^2=\frac{\mathbf{p}^2}{a^2}+m_n^2-\frac32 \frac{\ddot a}{a}-\frac34\Big(\frac{\dot a}{a}\Big)^2+\frac34 \Big(\frac{\dot W}{W}\Big)^2-\frac{\ddot W}{2W}}
can be solved iteratively.
In particular, when $\Delta m$ is sufficiently large that the heaviest mass (the mass just below $\Lambda_{\rm sp}$) in a tower   is much smaller than $\Lambda_{\rm sp}$, the leading term of $W$ near  $|\mathbf{p}|/a \sim \Lambda_{\rm sp}$ is just given by  $|\mathbf{p}|/a$.
\footnote{Since the physical momentum in the expanding universe is given by $\mathbf{p}/a$, the integration range of the momentum is taken to be $|\mathbf{p}|/a<\Lambda_{\rm sp}$, or equivalently, $|\mathbf{p}| <a\Lambda_{\rm sp}$.}
 Then to the first subleading order, we obtain \cite{Parker:2009uva} (see also \cite{Markkanen:2018bfx})
\dis{W^2\simeq \frac{\mathbf{p}^2}{a^2}+m_n^2-\frac16{\cal R}=\frac{\mathbf{p}^2}{a^2}+m_n^2-2H^2,}
where ${\cal R}$ is the Ricci scalar. 
The appearance of $-(1/6){\cal R}$ can be also found from the heat kernal expansion 
\cite{Parker:1984dj, Jack:1985mw}.
Since we are interested in the UV behavior of  the effective potential, i.e., the contributions from $|\mathbf{p}|/a\simeq \Lambda_{\rm sp}$ to the loop integral, we employ this as a good approximation of  $W$.

 
Then the one-loop effective potential generated by $\Phi_n$ is given by
\dis{V_{\rm eff}  & =-\frac{i}{2({\rm volume})} {\rm Tr}\log(\square +m_n(\phi)^2-2H^2)
\\
&=-\frac{i}{2a^3} \sum_{n=1}^{N_{\rm sp}(\phi)} \int^{a\Lambda_{\rm sp}(\phi)}\frac{dW d^3 p}{(2\pi )^4}{\rm Tr}\log(\square +m_n(\phi)^2-2H^2). \label{eq:Veffori}}
Here $1/a^3$ in the denominator reflects the fact that the physical volume is given by $a^3\times$(comoving volume).
After the Wick rotation $W\to iW$ and redefinition $p_4=a W$, we obtain
\dis{V_{\rm eff} =\frac{1}{2 a^4} \sum_{n=1}^{N_{\rm sp}(\phi)}\int_{p_E<a\Lambda_{\rm sp}(\phi)}\frac{ d^4 p_E}{(2\pi)^4} \log( p_E^2+a^2 (m_n(\phi)^2-2H^2)).}
The explicit calculation gives
{\small
\dis{V_{\rm eff}=-\frac{1}{64\pi^2}\sum_{n=1}^{N_{\rm sp}(\phi)}\Big[&\frac{\Lambda_{\rm sp}(\phi)^4}{2} - \Lambda_{\rm sp}(\phi)^4\log\Big( \Lambda_{\rm sp}(\phi)^2+m_n(\phi)^2-2H^2\Big)
- \Lambda_{\rm sp}(\phi)^2 (m_n(\phi)^2-2H^2)
\\
&+ (m_n(\phi)^2-2H^2)^2\log\Big(\frac{\Lambda_{\rm sp}(\phi)^2+m_n(\phi)^2-2H^2}{m_n(\phi)^2-2H^2}\Big)\Big],
}
}
expansion of which in terms of $\Lambda_{\rm sp}$ is written as
{\small
\dis{V_{\rm eff} =\sum_{n=1}^{N_{\rm sp}(\phi)}\Big[&-\frac{\Lambda_{\rm sp}(\phi)^4}{128\pi^2}+\frac{\Lambda_{\rm sp}(\phi)^4}{64\pi^2}\log\Big( \Lambda_{\rm sp}(\phi)^2 \Big)+\frac{\Lambda_{\rm sp}(\phi)^2}{32\pi^2}(m_n(\phi)^2-2H^2)
\\
&-\frac{(m_n(\phi)^2-2H^2)^2}{128\pi^2}-\frac{(m_n(\phi)^2-2H^2)^2}{64\pi^2}\log\Big(\frac{\Lambda_{\rm sp}(\phi)^2}{m_n(\phi)^2-2H^2}\Big)\Big].
}
}
Putting $m_n(\phi)=n\Delta m(\phi)$   and approximating the summation with the integral, one obtains  
{\small
\dis{ V_{\rm eff} =\Big[&-\frac{N_{\rm sp}(\phi)\Lambda_{\rm sp}(\phi)^4}{128\pi^2}+\frac{N_{\rm sp}(\phi)\Lambda_{\rm sp}(\phi)^4}{64\pi^2}\log\Big( \Lambda_{\rm sp}(\phi)^2 \Big)
\\
&+\frac{\Lambda_{\rm sp}(\phi)^2}{32\pi^2}
\Big(\frac{N_{\rm sp}(\phi)^3 \Delta m(\phi)^2}{3}-2N_{\rm sp}(\phi) H^2\Big)
\\
&-\frac{1}{128\pi^2}\Big[\frac{N_{\rm sp}(\phi)^5}{5}\Delta m(\phi)^4-\frac43 N_{\rm sp}(\phi)^3 H^2\Delta m(\phi)^2+ 4N_{\rm sp}(\phi) H^4\Big]
\\
&-\frac{1}{15\sqrt2 \pi^2}\frac{H^5}{\Delta m(\phi)}\log\Big(\frac{(N_{\rm sp}(\phi)\Delta m(\phi)-\sqrt2 H)(\Delta m(\phi)+\sqrt2 H)}{(N_{\rm sp}(\phi)\Delta m(\phi)+\sqrt2 H)(\Delta m(\phi)-\sqrt2 H)}\Big)
\\
&-\frac{1}{7200\pi^2}\Big(9N_{\rm sp}(\phi)^5 \Delta m(\phi)^4-70N_{\rm sp}(\phi)^3 H^2\Delta m(\phi)^2+480 N_{\rm sp}(\phi)H^4\Big)
\\
&-\frac{N_{\rm sp}(\phi)}{960\pi^2} (3N_{\rm sp}(\phi)^4 \Delta m(\phi)^4-20 N_{\rm sp}(\phi)^2 \Delta m(\phi)^2H^2+60 H^4)\log\Big(\frac{\Lambda_{\rm sp}(\phi)^2}{N_{\rm sp}(\phi)^2\Delta m(\phi)^2-2H^2}\Big).
\label{eq:Veffint1}}
}
Then from \eqref{eq:NspLsp}, $V_{\rm eff}$ can be rewritten as
{\small
\dis{V_{\rm eff} =&-\frac{\Delta m(\phi)^{2/3}m_{\rm Pl}^{10/3}}{128\pi^2}+\frac{\Delta m(\phi)^{2/3}m_{\rm Pl}^{10/3}}{64\pi^2}\log\Big(\Delta m(\phi)^{2/3}m_{\rm Pl}^{4/3} \Big)
\\
&+\frac{\Delta m(\phi)^{2/3}m_{\rm Pl}^{4/3}}{32\pi^2}\Big(\frac{m_{\rm Pl}^2}{3}-2H^2 \Big(\frac{m_{\rm Pl}}{\Delta m(\phi)}\Big)^{2/3} \Big)
\\
&-\frac{1}{128\pi^2}\Big[\frac{\Delta m(\phi)^{2/3}m_{\rm Pl}^{10/3}}{5} -\frac43 m_{\rm Pl}^2 H^2 + 4\Big(\frac{m_{\rm Pl}}{\Delta m(\phi)}\Big)^{2/3}  H^4\Big]
\\
&-\frac{1}{15\sqrt2 \pi^2}\frac{H^5}{\Delta m(\phi)}\log\Big(\frac{(\Delta m(\phi)^{1/3}m_{\rm Pl}^{2/3}-\sqrt2 H)(\Delta m(\phi)+\sqrt2 H)}{(\Delta m(\phi)^{1/3}m_{\rm Pl}^{2/3}+\sqrt2 H)(\Delta m(\phi)-\sqrt2 H)}\Big)
\\
&-\frac{m_{\rm Pl}^2}{7200\pi^2} \Big(9 \Delta m(\phi)^{2/3}m_{\rm Pl}^{4/3}-70 H^2+ 480 H^4\Big(\frac{m_{\rm Pl}}{\Delta m(\phi)}\Big)^{2/3} \Big)
\\
&-\frac{1}{960\pi^2}\Big(3  \Delta m(\phi)^{2/3}m_{\rm Pl}^{10/3}-20  m_{\rm Pl}^{2}H^2+60 \Big(\frac{m_{\rm Pl}}{\Delta m(\phi)}\Big)^{2/3} H^4\Big)\log\Big(\frac{\Delta m(\phi)^{2/3}m_{\rm Pl}^{4/3}}{\Delta m(\phi)^{2/3}m_{\rm Pl}^{4/3}-2H^2}\Big).
}
}
On the other hand, whereas $\Lambda_{\rm sp}=m_{\rm Pl}^{2/3}\Delta m(\phi)^{1/3}$ is required to be larger than $H$, the characteristic   scale  of dS space,  $\Delta m(\phi)$ is also larger than $H$ in many realistic cases.
The observed value of the cosmological constant in   the current universe gives  $H\sim 10^{-60}m_{\rm Pl}$, which is much smaller than the electroweak scale as well as the predicted energy scales of  new physics.
Moreover, it seems that the `effective theory of inflation', in which the curvature perturbation and the graviton are the low energy degrees of freedom below $H$, well describes the universe in the inflationary era \cite{Cheung:2007st, Weinberg:2008hq}.
Indeed, so far as we are interested in the model for dS space based on the four-dimensional particle description, the mass scale of a tower of states like the Kaluza-Klein modes or the string excitations is typically larger than $H$.
\footnote{In this case, approximations to obtain $V_{\rm eff}$ is valid only for $\Delta m(\varphi)=\Delta m(0){\rm exp}[-\frac{\varphi}{cm_{\rm Pl}}]$ much larger than $H$, thus $\varphi/(c m_{\rm Pl})$ much larger than $\log\frac{\Delta m(0)}{H}$ cannot be considered.}
This is also what the swampland conjectures concerning a tower of states   try to explain, as can be found in, for example, the discussion on the dark dimension  \cite{Anchordoqui:2023oqm}.
 When $m_{\rm Pl} \gg \Delta m(\phi) \gg H$, $V_{\rm eff}$ is dominated by
\dis{V_{\rm eff}&\simeq \frac{1}{4800 \pi^2}\Delta m(\phi)^{2/3}m_{\rm Pl}^{10/3}\Big(75 \log\Big(\frac{\Delta m(\phi)^{2/3}m_{\rm Pl}^{4/3}}{\Lambda_0^2}\Big)-1\Big)
\\
&=\frac{\Delta m(0)^{2/3} m_{\rm Pl}^{10/3}}{4800\pi^2}e^{-\frac23\frac{\varphi}{c m_{\rm Pl}}}\Big[-50\frac{\varphi}{c m_{\rm Pl}}+75 \log\Big(\frac{\Delta m(0)^{2/3}m_{\rm Pl}^{4/3}}{\Lambda_0^2}\Big)-1\Big],}
where in the second line $V_{\rm eff}$ is rewritten in terms of the canonically normalized field $\varphi$ defined in \eqref{eq:varphi}.
A scale    $\Lambda_0$ is introduced for renormalization.
We also note that there can be contributions to the vacuum energy from other fields, which do  not change under the evolution of $\varphi$.
Parametrizing them by a constant added to $V_{\rm eff}$, we arrive at
\dis{V_{\rm eff}&=-\frac{\Delta m(0)^{2/3} m_{\rm Pl}^{10/3}}{96\pi^2}e^{-\frac23\frac{\varphi}{c m_{\rm Pl}}} \Big(\frac{\varphi}{c m_{\rm Pl}}+A_B\Big)+B_B,
\\
&=-\frac{\Lambda_{\rm sp}(0)^2 m_{\rm Pl}^{2}}{96\pi^2}e^{-\frac23\frac{\varphi}{c m_{\rm Pl}}} \Big(\frac{\varphi}{c m_{\rm Pl}}+A_B\Big)+B_B
\\
&=-\frac{\Lambda_{\rm sp}(\varphi)^2 m_{\rm Pl}^{2}}{64\pi^2} \frac23\Big(\frac{\varphi}{c m_{\rm Pl}}+A_B\Big)+B_B\label{eq:Veffbos}}
where   \eqref{eq:NspLsp} is used  in the second line. 
Constants $A_B$ and $B_B$ are determined by the renormalization condition that the vacuum energy density at $\varphi=0$ is given by  $3m_{\rm Pl}^2 H^2$, which reads
\dis{-\frac{\Lambda_{\rm sp}(0)^2 m_{\rm Pl}^{2}}{96\pi^2}A_B+B_B=3m_{\rm Pl}^2 H^2.}
This does not fix the constants completely, allowing various behaviors of the potential as will be discussed in the next section.

On the other hand, for a fermionic  tower of states, an overall minus sign should be taken into account.
Moreover, the relations (see \cite{Parker:2009uva})
\dis{[\nabla_\mu, \nabla_\nu ]\psi=\frac14\gamma^\lambda \gamma^\sigma R_{\mu\nu\lambda \sigma}\psi,\quad\quad\gamma^\mu \gamma^\nu \gamma^\lambda \gamma^\sigma R_{\mu\nu\lambda\sigma}=-2{\cal R}}
give
\dis{\gamma^\mu \nabla(\gamma^\nu\nabla_\nu \psi)=\Big(\square -\frac14{\cal R}\Big)\psi=(\square-3 H^2)\psi,\label{eq:Diracsq}}
which replaces $2H^2$ in the effective potential generated by  a bosonic tower of states   by $3H^2$, but this  does not affect the leading terms of $V_{\rm eff}$.
 Since \eqref{eq:Diracsq} is the squared Dirac operator, $V_{\rm eff}\sim {\rm Tr}(\gamma^\mu \nabla_\mu+ i m_n)$ is given by $\sim \frac12{\rm Tr}(\square+m_n^2-3H^2)$.
 As an operator $\square+m_n^2-3H^2$ is diagonal to the spinor indices, the trace over the spinor indices for the Dirac spinor gives a factor 4, then $V_{\rm eff}$ becomes $\sim 4\times\frac12 {\rm tr}(\square+m_n^2-3H^2)$ where ${\rm tr}$ denotes the trace over the momentum space and the summation over the states in a tower below $\Lambda_{\rm sp}$.
 For the Weyl or Majorana spinor, the trace over the spinor indices   gives a factor 2, thus we obtain $V_{\rm eff} \sim 2\times\frac12 {\rm tr}(\square+m_n^2-3H^2)$.
 We can compare it with $V_{\rm eff}$ generated by one complex scalar, or equivalently, two real scalars given by $\sim -2\times\frac12 {\rm tr}(\square+m_n^2-2H^2)$ : both the Weyl spinor and one complex scalar have two real degrees of freedom, hence when they have the same mass and $H \to 0$, i.e., in the supersymmetric case, the sum of these one-loop effective potentials vanishes.
Therefore, we obtain
\dis{V_{\rm eff}=&f \frac{\Delta m(0)^{2/3} m_{\rm Pl}^{10/3}}{96\pi^2}e^{-\frac23\frac{\varphi}{c m_{\rm Pl}}}\Big( \frac{\varphi}{c m_{\rm Pl}}+A_F\Big)+B_F
\\
=&f \frac{\Lambda_{\rm sp}(0)^2 m_{\rm Pl}^{2}}{96\pi^2}e^{-\frac23\frac{\varphi}{c m_{\rm Pl}}}\Big( \frac{\varphi}{c m_{\rm Pl}}+A_F\Big)+B_F
\\
=&f\frac{\Lambda_{\rm sp}(\varphi)^2 m_{\rm Pl}^{2}}{64\pi^2} \frac23\Big( \frac{\varphi}{c m_{\rm Pl}}+A_F\Big)+B_F,\label{eq:Vefffer}}
where $f=4$ for the Dirac spinor and $f=2$ for the Weyl or Majorana spinor.
Here $V_{\rm eff}(\varphi=0)=3m_{\rm Pl}^2H^2$ is satisfied by requiring
\dis{f\frac{\Lambda_{\rm sp}(0)^2 m_{\rm Pl}^{2}}{96\pi^2}A_F+B_F=3m_{\rm Pl}^2H^2.}


We now consider the one-loop effective potential generated by    several towers of states.
In particular, we restrict our attention to the case in which  the canonically normalized modulus can be written in the simple form as \eqref{eq:varphi}, e.g., either one particular tower mass scale is much lighter than others or $n$ towers have the same mass scale.
Going back to \eqref{eq:Veffint1}, one finds that the dominant effective potential terms generated by several bosonic towers of states are given by
 \dis{V_{\rm eff}&=-\sum_i  \frac{\Lambda_{\rm sp}^4 N_i  }{96\pi^2}  \Big(\frac32 \log\Big(\frac{\Lambda_{\rm sp}^2}{\Lambda_0^2}\Big)+A_{B,i}\Big)+B_{B} 
 \\
 &=-\sum_i \frac{\Lambda_{\rm sp}^4 }{96\pi^2} \frac{\Lambda_{\rm sp}}{\Delta m_i}  \Big(\frac32 \log\Big(\frac{\Lambda_{\rm sp}^2}{\Lambda_0^2}\Big)+A_{B,i}\Big)+B_{B}, }
and the similar expression can be found for   fermionic towers of states with the extra minus sign attached. 
 When one of tower mass scales, say, $\Delta m_1$ is extremely light compared to others, $N_1$ is much lager than $N_i$ with $i=2,3,\cdots$, hence $\Lambda_{\rm sp}\simeq m_{\rm Pl}/N_1^{1/2}$.
 Then   the first line of above can be approximated by 
 \dis{V_{\rm eff}\simeq -\frac{m_{\rm Pl}^4  }{96\pi^2 N_1} \sum_i   \frac{N_i}{N_1} \Big(\frac32 \log\Big(\frac{\Lambda_{\rm sp}^2}{\Lambda_0^2}\Big)+A_{B,i}\Big)+B_{B}, }
 and since $N_i/N_1\ll 1$ for $i=2,3,\cdots$,   a tower corresponding to $i=1$ gives the dominant contribution to $V_{\rm eff}$. 
 Meanwhile, if $n$ towers have the same  and the lightest tower mass scale $\Delta m$,   \eqref{eq:manylambdaN} is satisfied,   then $V_{\rm eff}$ becomes
\dis{V_{\rm eff}&=-n\frac{\Delta m(0)^{2/3} m_{\rm Pl}^{10/3}}{n^{5/3} 96\pi^2}e^{-\frac23\frac{\varphi}{c m_{\rm Pl}}} \Big(\frac{\varphi}{c m_{\rm Pl}}+A_B\Big)+B_B,}
where $c=(\sum_i c_i)/n=c_i$ since contributions of all the bosonic towers of states to the wavefunction renormalization are the same.
If the effective  potential is generated by both the bosonic and fermionic towers of states of the same mass scale, 
\dis{n c= \sum_i c_i =  n_Bc_B+n_F c_F,}
where $n_B$ ($c_B$) and $n_F$ ($c_F$) are   the number (the loop factor in the wavefunction renormalization) of real bosonic and fermionic towers, respectively, with $n=n_B+n_F$, then $V_{\rm eff}$ is given by
\dis{V_{\rm eff}&=-(n_B- f n_F )\frac{\Delta m(0)^{2/3} m_{\rm Pl}^{10/3}}{n^{5/3} 96\pi^2}e^{-\frac23\frac{\varphi}{c m_{\rm Pl}}} \Big(\frac{\varphi}{c m_{\rm Pl}}+A \Big) +B.}
If $n_B= f n_F$ by supersymmetry (for example,  a chiral supermultiplet of ${\cal N}=1$ supersymmetry contains two real scalars and one Weyl spinor, hence we have $n_B=2$, $f=2$, and $n_F=1$), the leading term of $V_{\rm eff}$ is just given by the flat potential.

 We close this section with a comment on the renormalization condition on the vacuum energy, $V_{\rm eff}(\varphi=0)=3 m_{\rm Pl}^2 H^2$. 
 The strong version of the emergence proposal claims that the scalar potential is  an IR effect : the potential vanishes in UV and it is entirely generated by a tower of states which would be integrated out.
 This can be supported by an observation that the moduli  potentials   are typically generated by  fluxes, the background vacuum expectation values of the $p$-form gauge field strengths \cite{Castellano:2022bvr}.
 Since the gauge field strength plays a role of the kinetic term,  it vanishes in UV according to the emergence proposal, which leads to the vanishing UV flux potential.
 On the other hand, as can be found in the stabilization of the K\"ahler moduli, the flux potential can be combined with the non-perturbative effect and the uplift potential \cite{Kachru:2003aw, Balasubramanian:2005zx}.
 Whereas it is not clear that each of these additional ingredients is generated by integrating out a tower of states, even if it is the case, all these towers of states need not be associated with the single modulus $\varphi$.
 Some of moduli can be heavy enough such that their stabilized values are almost fixed, and integrating out them is reflected in the coefficients of the potential of $\varphi$, which are given by the renormalization conditions including $V_{\rm eff}(\varphi=0)=3 m_{\rm Pl}^2 H^2$.

 \begin{figure}[!t]
  \begin{center}
  \includegraphics[width=0.4\textwidth]{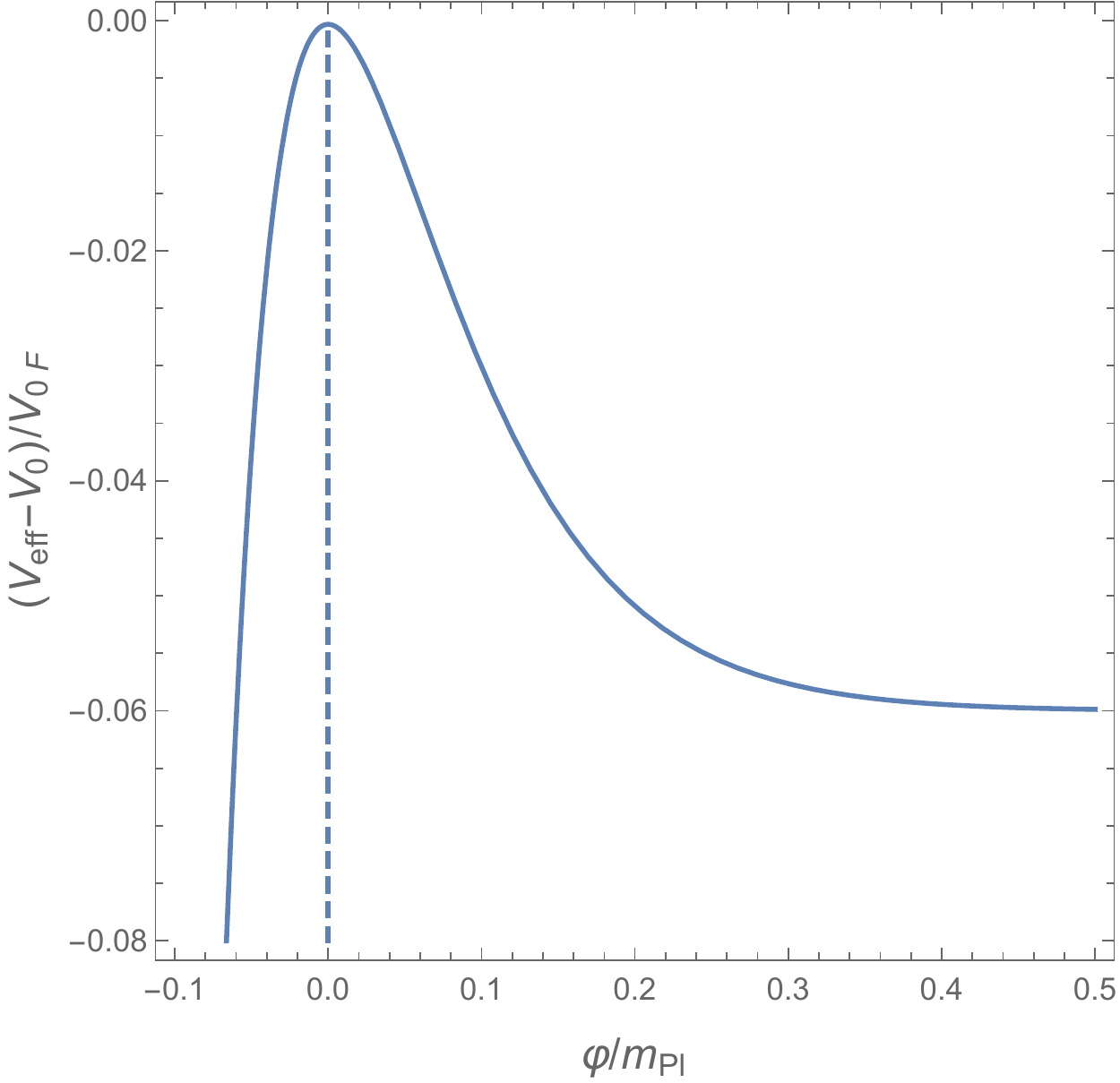}
  \end{center}
 \caption{The shape of $V_{\rm eff}(\varphi)$ generated by a fermionic tower  of states.
 Here the specific case in which $V_{\rm eff}$ is maximized at $\varphi=0$ is considered.
  The graph depicts the potential shifted below, $V_{\rm eff}(\varphi)-V_{\rm eff}(\varphi=0)$ such that a maximum value is given by zero. 
  }
\label{fig:VeffF}
\end{figure}

 \section{ de Sitter instability induced by a tower of states }
 \label{sec:dSstab}
 
As we have seen in the previous section, a descent of   a tower of states   from UV along the trajectory of the modulus $\varphi$ leads to the change in the vacuum energy, from which  $V_{\rm eff}(\varphi)$ is obtained. 
 This indicates that when the initial ($\varphi=0$) vacuum energy density is given by the positive value $3m_{\rm Pl}^2H^2$,  the (in)stability of dS space is determined by the shape of the potential which can be different depending on the choice of  constants $(A_B, B_B)$ or $(A_F, B_F)$ in \eqref{eq:Veffbos} and \eqref{eq:Vefffer},  respectively.

\subsection{Effects of a fermionic tower of sates}
 
  We first consider    $V_{\rm eff}(\varphi)$  generated by a fermionic tower of states in a loop,    given by \eqref{eq:Vefffer}.
 The extra minus sign of the closed fermion loop leads to the behavior of $V_{\rm eff}$ opposite to that generated by a bosonic tower of states.
 More concretely,  $V_{\rm eff}$ has a barrier separating    two minima, $V_{\rm eff}=-\infty$ at $\varphi=-\infty$ and $V_{\rm eff}=B_F$ at $\varphi=+\infty$.
  Suppose we choose  $(A_F, B_F)$ such that $V_{\rm eff}$ is maximized at  $\varphi=0$  as shown in Figure \ref{fig:VeffF}.
  Then the potential is given by
  \dis{V_{\rm eff}=f\frac{\Lambda_{\rm sp}(0)^2 m_{\rm Pl}^{2}}{96\pi^2}e^{-\frac23\frac{\varphi}{c m_{\rm Pl}}} \Big( \frac{\varphi}{c m_{\rm Pl}}+\frac32\Big)+\Big(-\frac32 f\frac{\Lambda_{\rm sp}(0)^2 m_{\rm Pl}^{2}}{96\pi^2} +3 m_{\rm Pl}^2 H^2\Big).}
  If $\varphi$ moves in the negative direction, the potential decreases indefinitely without bound, which is unphysical.
  Hence we restrict our discussion to the case in which $\varphi$ rolls down in the positive direction only.
  The transition from dS to AdS takes place, i.e., $V_{\rm eff}$ becomes zero, at
  \dis{ \frac23\frac{\varphi}{c m_{\rm Pl}}\simeq 2\pi\sqrt{\frac{96}{f}}\frac{H}{\Lambda_{\rm sp}(0)}.} 
  
 On the other hand, the slope of the potential  
  \dis{\frac{dV_{\rm eff}}{d\varphi}=-f\frac{\Lambda_{\rm sp}(0)^2  }{96\pi^2}e^{-\frac23\frac{\varphi}{c m_{\rm Pl}}}\times \frac{2}{3c^2}\varphi,}
  vanishes at $\varphi=0$ and $\varphi=\infty$ which correspond to the maximum and minimum of $V_{\rm eff}$, respectively.
  In the inflationary cosmology,   the decreasing rate   of    the vacuum energy along the inflaton trajectory is parametrized by the `potential slow-roll parameter' $\epsilon_V=(m_{\rm Pl}^2/2)[(dV/d\varphi)/V]^2$, which is well defined so far as $V$ is always positive (or always negative).
  In our case, however, $V_{\rm eff}$ can be either positive or negative, and in particular, $\epsilon_V$ defined by just replacing $V$ with $V_{\rm eff}$ diverges when  $V_{\rm eff}=0$.
 Therefore, we   parametrize the decreasing rate of the vacuum energy   by 
 \dis{\epsilon_V= \frac{m_{\rm Pl}^2}{2}\Big(\frac{dV_{\rm eff}/d\varphi}{V_{\rm eff}(\varphi)-V_{\rm eff}(\varphi=\infty)}\Big)^2,}
that is, we put $V_{\rm eff}(\varphi)-V_{\rm eff}(\varphi=\infty)$, the height of $V_{\rm eff}(\varphi)$ measured from the local minimum at $\varphi \to \infty$, instead of $V_{\rm eff}$ in the denominator.
An explicit calculation gives
\dis{\epsilon_V=\frac{2}{9 c^2}\Big(\frac{\varphi}{c m_{\rm Pl}}\Big)^2\frac{1}{\Big(\frac32+\frac{\varphi}{c m_{\rm Pl}}\Big)^2},}
which becomes one when 
\dis{\frac23\frac{\varphi}{c m_{\rm Pl}}=\frac{3 c}{\sqrt2-3c }\simeq 0.2,\label{eq:epsilonrange}}
where for the last estimation $c\simeq 1/(4\pi)$ is used.
 This value is larger than the value of $\varphi$ at which $V_{\rm eff}$ vanishes provided $H/\Lambda_{\rm sp}(0) <0.003 f^{1/2}$, which means that  the cosmological constant in this case becomes negative   even before $\epsilon_V$ becomes ${\cal O}(1)$.
 We also note that while $\epsilon_V$ is a monotonically increasing function of $\varphi$,  the increasing rate slows down  as $\varphi/(c m_{\rm Pl})$ becomes larger than $3/4$.
 
 Indeed, from
 \dis{\frac{d^2V_{\rm eff}}{d\varphi^2}=-\frac23 f\frac{\Lambda_{\rm sp}(0)^2}{96\pi^2 c^2}e^{-\frac23\frac{\varphi}{c m_{\rm Pl}}}\Big(1-\frac23\frac{\varphi}{c m_{\rm Pl}}\Big),}
which is negative and sizeable for $0<(2/3)[\varphi/(c m_{\rm Pl})]<1$,  one finds that $V_{\rm eff}$ rapidly decreases when $\varphi$ rolls down in this range.
This can be well quantified by defining 
   \dis{\eta_V=  m_{\rm Pl}^2  \frac{d^2V_{\rm eff}/d\varphi^2}{V_{\rm eff}(\varphi)-V_{\rm eff}(\varphi=\infty)} =-\frac{4}{9 c^2}  \frac{1-\frac23\frac{\varphi}{c m_{\rm Pl}}}{1+\frac23\frac{\varphi}{c m_{\rm Pl}}},}
   in the similar way to $\eta_V$ in the inflationary cosmology :
 when $(2/3)[\varphi/(c m_{\rm Pl})]$ moves from $0$ to $1$,  the absolute value $|\eta_V|$ decreases monotonically.
   In particular, since $\eta_V=-4/(9c^2)\sim -{\cal O}(1)$ at $\varphi=0$,  the vacuum energy quickly decreases at the initial stage. 
 Since the absolute value $|\eta_V|$ becomes smaller than one for
   \dis{\frac23\frac{\varphi}{c m_{\rm Pl}}>\frac{4-9c^2}{4+9c^2 }\simeq 0.97,\label{eq:etarange}}
   we can say that $\eta_V\sim -{\cal O}(1)$ in the region where $\epsilon_V$ is much smaller than one.

  It is remarkable that the behaviors of $V_{\rm eff}$ observed so far   are well consistent with the claims in the refined dS swampland conjecture \cite{Andriot:2018mav, Garg:2018reu, Ooguri:2018wrx}, which is supported by the distance conjecture  as well as the covariant entropy bound \cite{Ooguri:2018wrx}.
  That is, for small value of $\varphi$, $\epsilon_V<1$ but $\eta_V\sim -{\cal O}(1)$ so the positive $V_{\rm eff}$  decreases rapidly.
  Meanwhile,   as $\varphi/m_{\rm Pl}$ gets larger than ${\cal O}(1)$, $\epsilon_V$ becomes ${\cal O}(1)$. 
 On the other hand, one may realize the monotonically but slowly decreasing behavior of the potential by choosing $(A_F, B_F)$ such that
   \dis{V_{\rm eff}=f\frac{\Lambda_{\rm sp}(0)^2 m_{\rm Pl}^{2}}{96\pi^2}e^{-\frac23\frac{\varphi}{c m_{\rm Pl}}}   \frac{(\varphi+\varphi_0)}{c m_{\rm Pl}} +\Big(- f\frac{\Lambda_{\rm sp}(0)^2 m_{\rm Pl}^{2}}{96\pi^2}  \frac{ \varphi_0 }{c m_{\rm Pl}}+3 m_{\rm Pl}^2 H^2\Big),}
   where $\varphi_0/(cm_{\rm Pl})$ is taken to be much larger than one.
   Note that $\varphi_0/(cm_{\rm Pl})$ here is just a constant determined by the choice of $(A_F, B_F)$, not the specific field value.
   We can define $\epsilon_V$ and $\eta_V$ in the same way as the previous case, 
   \dis{\epsilon_V= \frac{m_{\rm Pl}^2}{2}\Big(\frac{dV_{\rm eff}/d\varphi}{V_{\rm eff}(\varphi)-V_{\rm eff}(\varphi=\infty)}\Big)^2,\quad\quad
  \eta_V=  m_{\rm Pl}^2  \frac{d^2V_{\rm eff}/d\varphi^2}{V_{\rm eff}(\varphi)-V_{\rm eff}(\varphi=\infty)}, \label{eq:epsetagen}}
an explicit calculation of which gives
\dis{&\epsilon_V=\frac{2}{9 c^2} \frac{\Big(\frac32-\frac{\varphi+\varphi_0}{c m_{\rm Pl}}\Big)^2}{\Big(\frac{\varphi+\varphi_0}{c m_{\rm Pl}}\Big)^2},
\\
&\eta_V=-\frac{4}{9 c^2}  \frac{1- \frac{\varphi+\varphi_0}{3c m_{\rm Pl}}}{ \frac{\varphi+\varphi_0}{3c m_{\rm Pl}}},}
respectively.
Then both   $\epsilon_V $ and $\eta_V $ are given by ${\cal O}(1)$, just like the quintessence model.

\subsection{Effects of a bosonic tower of sates}
 
 \begin{figure}[!t]
  \begin{center}
   \includegraphics[width=0.4\textwidth]{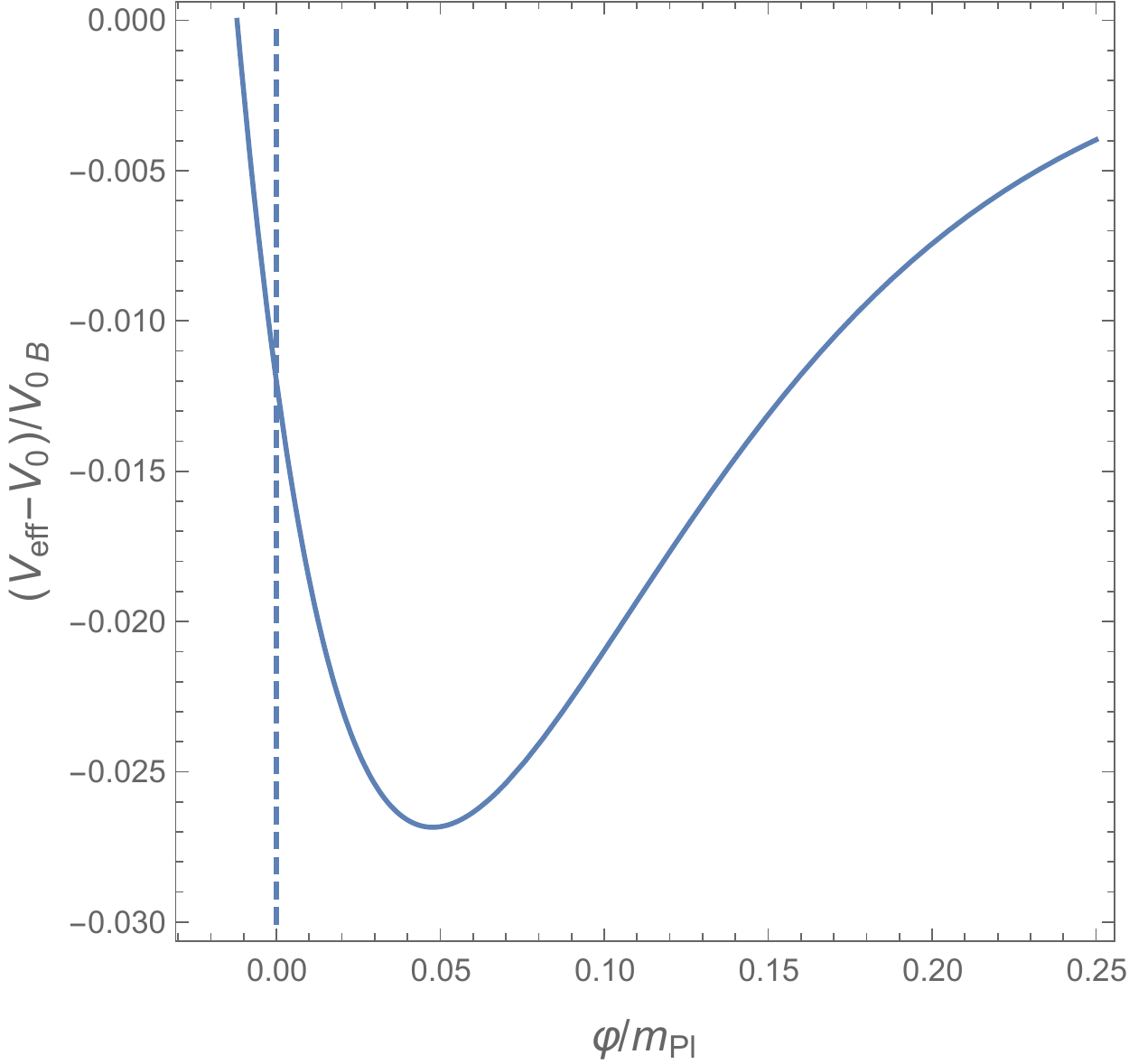}
  \end{center}
 \caption{The shape of $V_{\rm eff}(\varphi)$ generated by a bosonic tower  of states.
 The constants $(A_B, B_B)$ are chosen such that  $\varphi$ rolls down the potential.
  }
\label{fig:Veff}
\end{figure}

   We now consider the case in which  $V_{\rm eff}$ is generated by a bosonic tower of states in a loop.
  Requiring $V_{\rm eff}(\varphi=0)=3m_{\rm Pl}^2H^2$,  \eqref{eq:Veffbos} can be written in the form of
   \dis{V_{\rm eff}=-\frac{\Lambda_{\rm sp}(0)^2 m_{\rm Pl}^{2}}{96\pi^2}e^{-\frac23\frac{\varphi}{c m_{\rm Pl}}} \Big(\frac{ \varphi+\varphi_0 }{c m_{\rm Pl}}\Big) +\Big( \frac{\Lambda_{\rm sp}(0)^2 m_{\rm Pl}^{2}}{96\pi^2}  \frac{ \varphi_0 }{c m_{\rm Pl}}+3 m_{\rm Pl}^2 H^2\Big),}
   the shape of which is shown in Figure \ref{fig:Veff}.
  From 
  \dis{\frac{dV_{\rm eff}}{d\varphi}=-\frac{\Lambda_{\rm sp}(0)^2 m_{\rm Pl} }{96\pi^2 c} e^{-\frac23\frac{\varphi}{c m_{\rm Pl}}} \Big(1-\frac23 \frac{ \varphi+\varphi_0 }{c m_{\rm Pl}}\Big),}
 one finds that if $\varphi_0/(c m_{\rm Pl})<3/2$,  $\varphi$ initially at zero   rolls down   the potential to the minimum.
 On the other hand, when $\varphi_0/(c m_{\rm Pl})>3/2$, the vacuum energy increases as the value of $\varphi$ increases, hence $\varphi$ moves backward, i.e.,  in the negative direction until it is stabilized at the minimum, at which $\Delta m$ is heavier than $\Delta m(0)$.
 This can be one way to realize the stable dS vacuum : by choosing $\varphi_0/(c m_{\rm Pl})$ larger than but sufficiently close to $3/2$, the vacuum energy at the stabilized value of $\varphi$ can be still positive, and a bosonic tower of states no longer descends from UV.

In contrast, when $\varphi_0/(c m_{\rm Pl})<3/2$,  the vacuum energy decreases rapidly, which can be parametrized by    dimensionless parameters $\epsilon_V$ and $\eta_V$ defined by
   \dis{\epsilon_V= \frac{m_{\rm Pl}^2}{2}\Big(\frac{dV_{\rm eff}/d\varphi}{V_{\rm eff}(\varphi)-V_{\rm min}}\Big)^2,\quad\quad
  \eta_V=  m_{\rm Pl}^2  \frac{d^2V_{\rm eff}/d\varphi^2}{V_{\rm eff}(\varphi)-V_{\rm min}},}
 where in the denominator we put $V_{\rm eff}(\varphi)-V_{\rm min}$, the height of the potential measured from the minimum.
 Since $V_{\rm eff}$ is minimized at $\varphi=(3/2)cm_{\rm Pl}-\varphi_0$ and $V_{\rm min}$  is given by
 \dis{V_{\rm min}=-\frac{m_{\rm Pl}^2\Lambda_{\rm sp}(0)^2}{64\pi^2}e^{\frac23\frac{\varphi_0}{c m_{\rm Pl}}-1}\Big(1-\frac23\frac{\varphi_0}{c m_{\rm Pl}}e^{-\frac23\frac{\varphi_0}{c m_{\rm Pl}}+1}\Big)+3m_{\rm Pl}^2H^2,}
two parameters become
 \dis{&\epsilon_V=\frac{2}{9 c^2} \frac{\Big(\frac32-\frac{\varphi+\varphi_0}{c m_{\rm Pl}}\Big)^2}{\Big(\frac32 e^{\frac23\frac{\varphi+\varphi_0}{c m_{\rm Pl}}-1}-\frac{\varphi+\varphi_0}{c m_{\rm Pl}}\Big)^2},
\\
&\eta_V=\frac{4}{9 c^2}  \frac{3- \frac{\varphi+\varphi_0}{ c m_{\rm Pl}}}{ \frac32 e^{\frac23\frac{\varphi+\varphi_0}{c m_{\rm Pl}}-1}-\frac{\varphi+\varphi_0}{c m_{\rm Pl}}},}
respectively.
They indicate that in the limit of $\varphi_0/(c m_{\rm Pl})\ll 1$, both $\epsilon_V$ and $\eta_V$ are ${\cal O}(1)$ at $\varphi=0$ :  $\epsilon_V \simeq (2/9)(e/c)^2$ and $\eta_V\simeq (8/9)(e/c^2)$, respectively.
In particular,  $\eta_V$ is positive for $\varphi<3 c m_{\rm Pl}-\varphi_0$ hence $\varphi$ moves very quickly at the initial stage where the vacuum energy is positive.
Indeed, when  $\varphi_0/(c m_{\rm Pl})\ll 1$, the vacuum energy becomes zero at
\dis{\frac{\varphi}{c m_{\rm Pl}}\simeq 3\times 96\pi^2\Big(\frac{H}{\Lambda_{\rm sp}(0)}\Big)^2\frac{1}{1-\frac23\frac{ \varphi_0}{c m_{\rm Pl}}}.}

On the other hand,  when $\varphi_0/(c m_{\rm Pl})$ is close to $3/2$, the initial value $\varphi=0$ is near to the minimum of the potential which is still positive, hence $V_{\rm eff}$ will be stabilized at the dS minimum even before $\Delta m$  sufficiently decreases.
Therefore, unlike the effective potential generated by a fermionic tower of states in a loop, that generated by a bosonic tower of states in a   loop seems to allow the stable dS vacuum at which a bosonic tower of states is still decoupled from the EFT.
This implies that the distance conjecture alone is not sufficient to show the instability of the dS vacuum at least at the field theoretic level.
This may not be surprising since in the argument of \cite{Ooguri:2018wrx} supporting the dS swampland conjecture, another conjecture of the covariant entropy bound is imposed in addition to the distance conjecture.
 In order to argue the dS instability at the field theoretic level, we need to explain how the contributions from a bosonic tower of states can be suppressed at initial stage where $\varphi$ is much smaller than $c m_{\rm Pl}$ and the vacuum energy is positive.
 Presumably, supersymmetry can be an additional ingredient for the dS instability since even if supersymmetry is imposed in UV, it must be broken to realize dS space.
 Such a supersymmetry breaking typically makes the bosonic mass scale much heavier than the mass scale of the fermionic partner, so the mass scale of a fermionic tower of states can be smaller than that of a bosonic tower of states.
 Then the total effective potential can be dominated by   the contribution from a fermionic tower of states in a loop, at least at initial stage where supersymmetry breaking is strongest. 
 Moreover, whereas $V_{\rm eff}$ generated by a fermionic tower of states in a loop decreases indefinitely for $\varphi<0$, this can be cancelled by the   loop contribution from the bosonic superpartner, which is possible in the presence of supersymmetry.
 Indeed, in the region of the negative potential, supersymmetry is restored, thus the bosonic and fermionic loop contributions become the same in size but   opposite in sign.
 Then we expect other effects like the non-perturbative term must be dominant to stabilize $\varphi$ (for relevant discussion based on the string model, see, e.g., \cite{Andriot:2021rdy}).
 
 The fact that the stabilization of $\varphi$ requires the sum of several potential terms may imply that it is difficult to relate the vacuum energy density at the stabilized value of $\varphi$ to the tower mass scale, as the  AdS/dS distance conjecture \cite{Lust:2019zwm}  claims. 
 Indeed, our analysis can be applied to the rolling behavior of $\varphi$ in the region where one specific one-loop effective potential generated by a tower of states is dominant, rather than the stabilization of $\varphi$.
 Nevertheless,  the one-loop effective potential  generated by a tower of states becomes exponentially small ($V_{\rm eff} \sim \Delta m(\varphi)^{2/3}$) for the trans-Planckian value of $\varphi$ at which the tower mass scale becomes light thus  the effective field theory is no longer valid. 
 In other words, the part of the potential, instead of the vacuum energy density at the stabilized value, becomes small as the tower mass scale gets light along the increasing value of $\varphi$. 
 The similar situation can be found in the string model for the metastable dS vacuum.
 When the uplift potential is generated by $\overline{{\rm D}3}$-branes at the tip of the throat \cite{Kachru:2002gs}, the uplift potential is redshifted in the same way as the Kaluza-Klein mass scale, satisfying the scaling behavior $V_{\rm uplift} \sim m_{\rm KK}^4$ \cite{Blumenhagen:2022zzw, Seo:2023ssl}.

 \section{Conclusions}
\label{sec:conclusion}

In this article, we investigate the (in)stability of dS space by assuming the distance conjecture and the strong version of the emergence proposal such that  both the effective potential and the kinetic term of the modulus are generated by integrating out a tower of states.  
In particular,  we focus on the simple cases, in which one  particular tower mass scale  is extremely light or several towers of states have the same mass scale.
Then    the canonically normalized modulus is given by the exponent of the tower mass scale.
 When a tower of states is fermionic, the one-loop effective potential  is more or less consistent with the dS swampland conjecture: either $m_{\rm Pl}|\nabla V/V|\sim{\cal O}(1)$ or $m_{\rm pl}^2 \nabla^2 V/V\sim -{\cal O}(1)$ is satisfied.  
 In contrast, because of the  absence of the extra minus sign in the closed loop, the effective potential generated by a bosonic tower of states shows the opposite behavior : the modulus can be stabilized in the dS vacuum. 
  Therefore, at least at the field theoretic level, the instability of dS space requires that at the early stage of the traverse of the modulus, the one-loop effective potential is dominated by the loop contributions from a fermionic tower of states.
  In other words, when the vacuum energy is positive,    the fermionic tower mass scale needs to be much lighter than the bosonic one.
 This can be  achieved when supersymmetry is  broken since in this case the vacuum energy can be positive and  the bosonic mass scale is heavier than the fermionic mass scale.  
 As another (and presumably equivalent) possibility, since the dS swampland conjecture is also supported by the covariant entropy bound,  the thermodynamic properties of quantum gravity may provide supplementary reason to argue the instability of dS space (for relevant discussions, see, e.g., \cite{Seo:2019mfk, Seo:2019wsh, Cai:2019dzj, Aalsma:2019rpt, Gong:2020mbn, Blumenhagen:2020doa, Seo:2021bpb, Castellano:2021mmx, Seo:2022uaz}).

\subsection*{Acknowledgements}

 MS is grateful to an anonymous referee for a number of suggestions which improved the manuscript significantly.

%

%


\appendix

\renewcommand{\theequation}{\Alph{section}.\arabic{equation}}

\section{A tower of states associated with two tower mass scales}
\label{app:multitower}
\setcounter{equation}{0}

 In this appendix, we discuss how the expressions for $N_{\rm sp}$ and $\Lambda_{\rm sp}$ are modified when a tower of states is associated with   two tower mass scales.
 Our study also shows that the simple expression for the canonically normalized field like \eqref{eq:varphi} is not  defined in this case.
 
 In the presence of two tower mass scales $\Delta m_1$ and $\Delta m_2$, the squared mass of the state in a tower is given by
 \dis{m_{\mathbf n}^2= n_1^2 \Delta m_1^2 + n_2^2 \Delta m_2^2,\label{eq:multitower}}
with $n_1$ and $n_2$ integers. 
If both $\Delta m_1$ and $\Delta m_2$ are sufficiently smaller than $\Lambda_{\rm sp}$, the number of states below $\Lambda_{\rm sp}$ is proportional to $1/4$ of the area of the ellipse with semi-axes $N_1=\Lambda_{\rm sp}/\Delta m_1$ and $N_2=\Lambda_{\rm sp}/\Delta m_2$ : $N_{\rm sp}\simeq N_1 N_2 = \Lambda_{\rm sp}^2/(\Delta m_1 \Delta m_2)$ with ${\cal O}(1)$ coefficient   omitted.
Since the species scale is still given by $\Lambda_{\rm sp}=m_{\rm Pl}/\sqrt{N_{\rm sp}}$, we obtain
\dis{N_{\rm sp}=\frac{m_{\rm Pl}}{\Delta m_1^{1/2} \Delta m_2^{1/2}},\quad\quad \Lambda_{\rm sp}=\Delta m_1^{1/4} \Delta m_2^{1/4}m_{\rm Pl}^{1/2}.}
Indeed, when  a tower of states is  associated with  $N$ tower mass scales labelled by $\Delta m_i$, $N_{\rm sp}$ and $\Lambda_{\rm sp}$ are given by
\dis{N_{\rm sp}=\frac{m_{\rm Pl}^{\frac{2N}{N+2}}}{\prod_{i=1}^N\Delta m_i^{\frac{2}{N+2}}},\quad\quad \Lambda_{\rm sp}=m_{\rm Pl}^{\frac{2}{N+2}}\prod_{i=1}^N\Delta m_i^{\frac{1}{N+2}},}
respectively, which are consistent with above as well as \eqref{eq:NspLsp}.

 Now suppose both $\Delta m_1$ and $\Delta m_2$ are determined by the single modulus $\phi$ (if only one of tower mass scales, say,  $\Delta m_1$, is determined by $\phi$, we just take $\partial_\phi \Delta m_2=0$).
 Then the contribution of a bosonic tower of states to the wavefunction renormalization of $\phi$ is proportional to
 \dis{Z_{\phi\phi}\propto \sum_{n_1}\sum_{n_2}(\partial_\phi m_{\mathbf n})^2,\label{eq:multiZ}}
 where the summation is taken over the values of $n_1$ and $n_2$ satisfying $m_{\mathbf n} < \Lambda_{\rm sp}$.
 From \eqref{eq:multitower}, one finds that
 \dis{(\partial_\phi m_{\mathbf n})^2=n_1^4\Big(\frac{\Delta m_1}{m_{\mathbf n}}\Big)^2 (\partial_\phi\Delta m_1)^2+ n_2^4\Big(\frac{\Delta m_2}{m_{\mathbf n}}\Big)^2 (\partial_\phi\Delta m_2)^2 + 2 n_1^2n_2^2 \frac{\Delta m_1}{m_{\mathbf n}}\frac{\Delta m_2}{m_{\mathbf n}} \partial_\phi\Delta m_1 \partial_\phi\Delta m_2.}
 Since $m_{\mathbf n}$ depends on both $n_1$ and $n_2$,   the RHS of \eqref{eq:multiZ} can be approximated by
 \dis{\sum_{n_1}\sum_{n_2}(\partial_\phi m_{\mathbf n})^2=&I_1(\Delta m_1, \Delta m_2)(\partial_\phi\Delta m_1)^2+I_1(\Delta m_2, \Delta m_1)(\partial_\phi\Delta m_2)^2 
 \\
 &+ 2I_2(\Delta m_1, \Delta m_2)\partial_\phi\Delta m_1 \partial_\phi\Delta m_2, }
 where
 \dis{&I_1 (\Delta m_1, \Delta m_2)=\int d n_1 dn_2\frac{n_1^4 \Delta m_1^2}{n_1^2 \Delta m_1^2+n_2^2\Delta m_2^2},
 \\
 &I_2(\Delta m_1, \Delta m_2)=\int d n_1 dn_2 \frac{n_1^2 n_2^2 \Delta m_1\Delta m_2}{n_1^2 \Delta m_1^2+n_2^2\Delta m_2^2}. }
 These integrals can be evaluated by changing the integration variables from $(n_1, n_2)$ to $(r, \theta)$ defined by $n_1=(r/\Delta m_1)\cos\theta$ and $n_2=(r/\Delta m_2)\sin\theta$.
 Since $r$ and $\theta$ lie in the range $0 \leq r < \Lambda_{\rm sp}$ and $0\leq \theta \leq \pi/4$, respectively,  we obtain
\dis{I_1 (\Delta m_1, \Delta m_2)=\frac{3\pi}{64}\frac{m_{\rm Pl}^2}{\Delta m_1^2},\quad\quad
I_2(\Delta m_1, \Delta m_2)=\frac{\pi}{64}\frac{m_{\rm Pl}^2}{\Delta m_1 \Delta m_2}.}
Therefore, the RHS of \eqref{eq:multiZ} is given by
\dis{\sum_{n_1}\sum_{n_2}(\partial_\phi m_{\mathbf n})^2 = \frac{3\pi}{64}m_{\rm Pl}^2\Big[\Big(\frac{\partial_\phi\Delta m_1}{\Delta m_1}\Big)^2+ \Big(\frac{\partial_\phi\Delta m_2}{\Delta m_2}\Big)^2+\frac23 \frac{\partial_\phi\Delta m_1}{\Delta m_1}\frac{\partial_\phi\Delta m_2}{\Delta m_2}\Big],}
which cannot be written in the form of $[\partial_\phi F(\Delta m_1, \Delta m_2)]^2$ for some function $F(\Delta m_1, \Delta m_2)$ unless one of $\partial_\phi \Delta m_1$ and $\partial_\phi \Delta m_2$ vanishes.
Therefore,  $Z_{\phi\phi}(\partial_\mu\phi)^2$ cannot be written in the form of $(\partial_\mu \varphi)^2$.
We note that such a failure of obtaining the simple canonically normalized field originates from the fact that the numerical coefficient of $I_1$ ($3\pi/64$) is different from that of $I_2$ ($\pi/64$).
More precisely,  $I_1$ and $I_2$ have the same $r$ integration, but different  $\theta$ integrations given by
\dis{\int_0^{\pi/2}\cos^4\theta d\theta =\frac{3\pi}{16},\quad\quad \int_0^{\pi/2}\cos^2\theta \sin^2\theta d\theta=\frac{\pi}{16},\label{eq:cossin}
}
respectively.
The similar analysis also shows that  we cannot rewrite $Z_{\phi\phi}$ generated by a fermionic tower of states,
\dis{Z_{\phi\phi}\propto \sum_{n_1}\sum_{n_2}(\partial_\phi m_{\mathbf n})^2\log\Big(\frac{\Lambda_{\rm sp}^2}{m_{\mathbf n}^2}\Big)\label{eq:multiZphifer}}
 in a simple form, $[\partial_\phi F(\Delta m_1, \Delta m_2)]^2$.
 To see this, we note that the RHS is estimated as
 \dis{ \sum_{n_1}\sum_{n_2}(\partial_\phi m_{\mathbf n})^2\log\Big(\frac{\Lambda_{\rm sp}^2}{m_{\mathbf n}^2}\Big)  \simeq &I'_1(\Delta m_1, \Delta m_2)(\partial_\phi\Delta m_1)^2+I'_1(\Delta m_2, \Delta m_1)(\partial_\phi\Delta m_2)^2 
 \\
 &+ 2I'_2(\Delta m_1, \Delta m_2)\partial_\phi\Delta m_1 \partial_\phi\Delta m_2,}
 where
  \dis{&I'_1 (\Delta m_1, \Delta m_2)=\int d n_1 dn_2\frac{n_1^4 \Delta m_1^2}{n_1^2 \Delta m_1^2+n_2^2\Delta m_2^2}\log\Big(\frac{\Lambda_{\rm sp}^2}{n_1^2 \Delta m_1^2+n_2^2\Delta m_2^2}\Big),
 \\
 &I'_2(\Delta m_1, \Delta m_2)=\int d n_1 dn_2 \frac{n_1^2 n_2^2 \Delta m_1\Delta m_2}{n_1^2 \Delta m_1^2+n_2^2\Delta m_2^2}\log\Big(\frac{\Lambda_{\rm sp}^2}{n_1^2 \Delta m_1^2+n_2^2\Delta m_2^2}\Big). } 
 In these integrals,  $\log(\Lambda_{\rm sp}^2/m_{\mathbf n}^2)=2\log(\Lambda_{\rm sp}/r)$ depends only on $r$ so the $r$ integration in $I'_1$ is  the same as that in $I'_2$, but the $\theta$ integrations in $I'_1$ and $I'_2$ are still given by \eqref{eq:cossin}.
  Therefore, the RHS of  \eqref{eq:multiZphifer} is written in the form of
  \dis{\sum_{n_1}\sum_{n_2}(\partial_\phi m_{\mathbf n})^2 \log\Big(\frac{\Lambda_{\rm sp}^2}{m_{\mathbf n}^2}\Big)=  3 k m_{\rm Pl}^2\Big[\Big(\frac{\partial_\phi\Delta m_1}{\Delta m_1}\Big)^2+ \Big(\frac{\partial_\phi\Delta m_2}{\Delta m_2}\Big)^2+\frac23 \frac{\partial_\phi\Delta m_1}{\Delta m_1}\frac{\partial_\phi\Delta m_2}{\Delta m_2}\Big],}
and the explicit calculation fixes the value of the coefficient  $k$ by $\pi/128$.
In any case, $Z_{\phi\phi}(\partial_\mu\phi)^2$ cannot be written in the form of $(\partial_\mu \varphi)^2$ even if $Z_{\phi\phi}$ is generated by a fermionic tower of states.

\end{document}